\documentclass[%
 aip,
 amsmath,amssymb,
 reprint,%
]{revtex4-2}

\usepackage{graphicx}
\usepackage{dcolumn}
\usepackage{bm}

\usepackage[utf8]{inputenc}
\usepackage[T1]{fontenc}
\usepackage{mathptmx}

\begin{document}

\preprint{AIP/123-QED}

\title[]{Early Detection of Thermoacoustic Instabilities in a Cryogenic Rocket Thrust Chamber using Combustion Noise Features and Machine Learning}

\author{Günther Waxenegger-Wilfing}
 \email{guenther.waxenegger@dlr.de}
\affiliation{Institute of Space Propulsion, German Aerospace Center (DLR), 74239 Hardthausen, Germany}
\author{Ushnish Sengupta}
\affiliation{Department of Engineering, University of Cambridge, Cambridge, Cambridgeshire CB2 1PZ, United Kingdom}
\author{Jan Martin}
\author{Wolfgang Armbruster}
\author{Justin Hardi}
\affiliation{Institute of Space Propulsion, German Aerospace Center (DLR), 74239 Hardthausen, Germany}
\author{Matthew Juniper}
\affiliation{Department of Engineering, University of Cambridge, Cambridge, Cambridgeshire CB2 1PZ, United Kingdom}
\author{Michael Oschwald}
 \altaffiliation[Also at ]{Institute of Jet Propulsion and Turbomachinery, RWTH Aachen University, 52062 Aachen, Germany}
\affiliation{Institute of Space Propulsion, German Aerospace Center (DLR), 74239 Hardthausen, Germany}

\date{\today}

\begin{abstract}
Combustion instabilities are particularly problematic for rocket thrust chambers because of their high energy release rates and their operation close to the structural limits. In the last decades, progress has been made in predicting high amplitude combustion instabilities but still, no reliable prediction ability is given. Reliable early warning signals are the main requirement for active combustion control systems. In this paper, we present a data-driven method for the early detection of thermoacoustic instabilities. Recurrence quantification analysis is used to calculate characteristic combustion features from short-length time series of dynamic pressure sensor data. Features like the recurrence rate are used to train support vector machines to detect the onset of an instability a few hundred milliseconds in advance. The performance of the proposed method is investigated on experimental data from a representative LOX/H$_2$ research thrust chamber. In most cases, the method is able to timely predict two types of thermoacoustic instabilities on test data not used for training. The results are compared with state-of-the-art early warning indicators.
\end{abstract}

\maketitle

\begin{quotation}
High-amplitude pressure oscillations known as combustion instability are an issue for all chemical propulsion systems. Combustion instabilities are particularly problematic for rocket thrust chambers because of their high energy release rates. Especially thermoacoustic oscillations are a major hazard, but difficult to predict. An important question is whether features of combustion noise can be used to construct reliable early warning signals for representative rocket thrust chambers. Among other things, instability precursors are needed for active combustion control systems. In this study, we use recurrence quantification analysis and well-known machine learning models to construct an early warning signal that should be able to predict high-amplitude instabilities a few hundred milliseconds in advance. The performance of the proposed method is investigated on experimental data from a representative LOX/H$_2$ research thrust chamber. We describe the test case in detail and also evaluate other measures like the Hurst exponent.
\end{quotation}

\section{Introduction}
\label{sec_intro}

Future space missions and their increasing complexity require rocket engines with the capability of adjustment to the specific missions and therefore a wide envelope of operating conditions \cite{colas2019}.
Today rocket engines are built and tested in terms of operational stability for the operating conditions they are explicitly built for. Due to the complexity of upcoming space missions reusing an engine without extensive testing for the new requirements which have to be met can result in a decisive advantage.

The occurrence of combustion instabilities presents high technical risks during rocket engine development projects and their operation \cite{Harrje}. Combustion instability refers to high-amplitude pressure oscillations during a combustion process and is an issue for all chemical propulsion systems, e.g. gas turbines. Combustion instabilities are particularly problematic for rocket thrust chambers due to their operation close to the structural limits caused by the necessity of extreme lightweight construction \cite{Heister}. Those instabilities can be further sub-categorized into high-frequency (screeching) and low-frequency instabilities \cite{Harrje,Yang}. 
In the latter case, a further distinction in chugging and pogo instabilities is applied. Pogo is characterized by a change of the external loads (e.g. thrust modulations) which are transferred to the structure and chugging is normally attributed to an interaction between the combustion process and
propellant feed system elasticity. In the case of high-frequency instabilities, the frequency of pressure oscillations is typically linked to the acoustic modes of the combustion chamber. The existing feedback loop within these acoustic oscillations was first described by Lord Rayleigh \cite{Rayleigh1878}. If the pressure oscillation and the heat release are in phase the necessary condition for growing amplitudes is given. Otherwise, a damping effect applies. Due to the enormous power density in combustion devices for rockets, the transfer of a small amount of the total heat release rate into the acoustic field can cause rapidly growing pressure oscillations to damagingly high amplitudes. In the last decades, progress has been made in predicting high amplitude combustion instabilities but still, no absolute certainty has yet been achieved \cite{Schulze17}.

Two different approaches can be used to prevent the occurrence of thermoacoustic instability.  On the one hand, the development of these oscillations can be suppressed via passive control (for example baffles or acoustic liners \cite{Harrje}) or on the other hand through active control. In the case of active control, the system has to be able to process the data in time by using available sensor time-series data. Questions regarding suitable sensors, actuators, and optimal control targets must also be clarified  Furthermore, high robustness and reliability are mandatory. Due to the challenging requirements for an active control system no usage of such a system for an engine in operational service is known to the authors.

The construction of instability precursors from pressure measurements has been intensively studied by the thermoacoustic instability community \cite{Juniper2018}. The first step in this direction was taken by Lieuwen \cite{Lieuwen2005}, who used the autocorrelation decay of combustion noise filtered around an acoustic eigenfreuqency to obtain an effective damping coefficient that indicated closeness to instability. Other researchers have borrowed measures from nonlinear time series analysis which capture the transition from the chaotic behaviour displayed by stable turbulent combustors to the deterministic acoustics during instability, e.g. Gottwald's 0-1 test \cite{Nair2013}. Similarly, Gotoda and coworkers have used the Wayland test for nonlinear determinism \cite{Gotoda2011} to assess the stability margin of a combustor. Nair et al \cite{Nair2014intermittency} reported that instability is often presaged by intermittent bursts of high-amplitude oscillations and used recurrence quantification analysis to detect these. In a later paper \cite{Nair2014}, Nair and coworkers noted that the multifractality of combustion noise tends to disappear as the system transitions to instability and suggested that a decline in the Hurst exponent is a good indicator of this phenomenon. Further studies have demonstrated that measures derived from symbolic time series analysis (STSA) \cite{sarkar2016dynamic} and complex networks \cite{murugesan2016detecting} are also able to capture the onset of instability.

Machine learning tools have been employed to extract information from pressure signals instead of relying on hand-crafted precursors of instability. Hidden Markov models constructed from the output of STSA \cite{jha2018markov} or directly from pressure measurements \cite{mondal2018early} have been used to classify the state of combustors. Hachijo et al \cite{Hachijo2019} have projected pressure time series onto the entropy-complexity plane and used support vector machines (SVMs) to predict thermoacoustic instability. SVMs were also employed by Kobayashi et al \cite{Kobayashi2019} who used them in combination with principal component analysis and ordinal pattern transition networks to build precursors from simultaneous pressure and chemiluminiscence measurements. Sengupta et al \cite{sengupta2020bayesian} show that the power spectrum of the noise can be used to predict the linear stability of a thermoacoustic eigenmode using Bayesian neural networks. Related work by McCartney et al \cite{mccartney2020} uses the detrended fluctuation analysis (DFA) spectrum of the pressure signal as input to a random forest and finds that this approach compares favorably to precursors from the literature. Recent works have investigated machine learning methods for the design and operation of cryogenic rocket engines \cite{Dresia2019,Waxenegger-Wilfing2020,Waxenegger-Wilfing2020a,Waxenegger-Wilfing2020b}.

In this paper, we investigate the question of whether the features of combustion noise are sufficient to reliably predict the occurrence of instabilities in a cryogenic rocket thrust chamber. Certain combustion noise characteristics are believed to be widely unaffected by the geometrical boundary conditions of an engine and thus enable applicability to a wide range of combustion devices with little or no adaption.

Our main contributions are the following:
\begin{itemize}
    \item development of a data-driven method to construct early warning signals for thermoacoustic instabilities using nonlinear combustion noise features and support vector machines
    \item quantitative evaluation of the constructed early warning signals on data from a representative cryogenic rocket thrust chamber
    \item comparison with state-of-the-art early warning indicators
\end{itemize}

The remainder of the paper is structured as follows: Section \ref{sec_bkd} describes the research thrust chamber "D" (BKD) and the experimental data. The basics of recurrence quantification analysis are outlined in section \ref{sec_rqa}. Section \ref{sec_svm} discusses support vector machines. Section \ref{sec_method} and section \ref{sec_test_case} present the data-driven method for construction of early warning signals and the test case respectively. The results, including the comparison with the performance of other early warning indicators, are shown in section \ref{sec_results}. Finally, section \ref{sec_conclusion} provides concluding remarks.

\section{Rocket Thrust Chamber "BKD"}
\label{sec_bkd}

This study is performed with experimental data of the research thrust chamber "D" (BKD), which is operated at the European Research and Technology test bench P8 at the DLR Institute of Space Propulsion in Lampoldshausen. BKD has been designed for heat transfer research with regenerative cooling of liquid hydrogen (LH$_2$) at representative conditions of European industrial LOX/H$_2$ engines, such as Vulcain or Vinci \cite{Sender,Hardi18_AA}. However, the thrust chamber showed self-excited high-frequency combustion instabilities during the first tests for LOX/H$_2$ combustion. For that reason, BKD has become a valuable platform for the analysis of chamber acoustic phenomena and the underlying coupling mechanism due to the representative geometry and conditions \cite{SGDiss,SGJPP,SGCEAS,SGPropPhys,Hardi18_AA,Armbruster18_AA,Armbruster19JPP}. In later tests, BKD was also used with a similar setup for the investigation of LOX/LNG combustion, which also showed high-frequency combustion instabilities \cite{Martin2020ST}. However, the current study only focuses on the prediction of thermoacoustic instabilities for the propellant combination LOX/H$_2$ in BKD.
\begin{figure}
\includegraphics{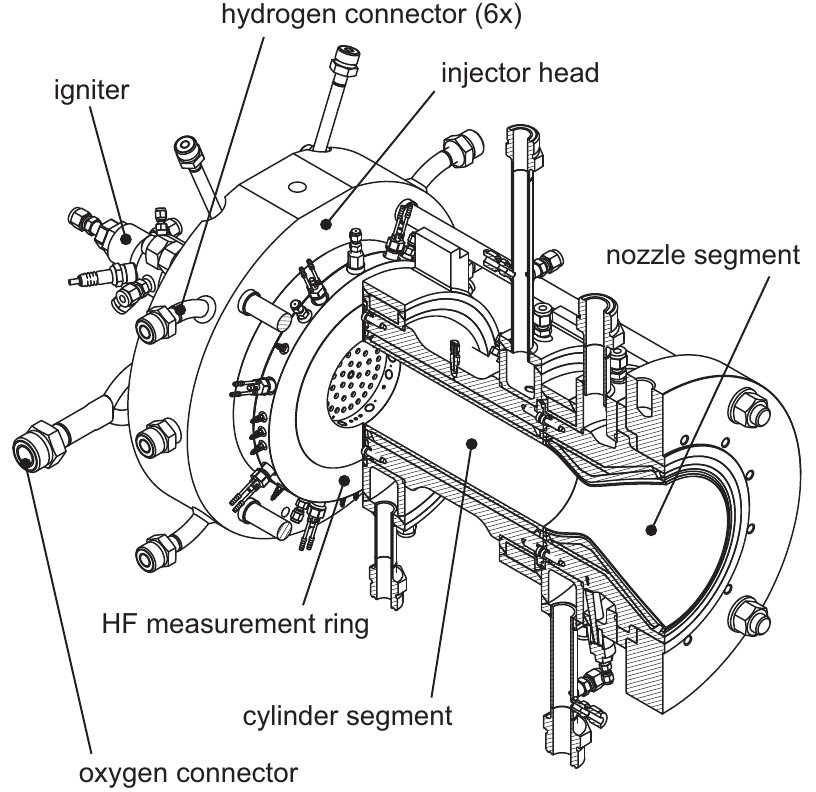}
\caption{\label{fig_BKD} Experimental thrust chamber BKD \cite{SGJPP}.}
\end{figure}

The BKD setup is shown in Fig. \ref{fig_BKD}. It consists of three main components: a multi-element injector head, a cylindrical combustion chamber, and a convergent-divergent nozzle. The injector head has 42 shear coaxial injection elements. The cylindrical combustion chamber has an inner diameter of 80~mm and is 200~mm long. For the instability investigations, the combustion chamber is water-cooled in order to guarantee a sufficient safety margin with respect to increasing thermal loads. The nozzle has a throat diameter of 50~mm, which leads to representative chamber characteristics as a contraction ratio of $\epsilon_c=2.56$ and a characteristic chamber length of $L^*=0.64$~m.

For the investigation of the stability behavior of this particular device, a large number of different operating conditions have been tested in several test campaigns. An operating condition or load point (LP) is thereby primarily defined by the mean chamber pressure $p_{cc}$, the propellant mixture ratio (ROF = $\dot m_{O2}/\dot m_{H2}$) and the propellant injection temperatures ($T_{O2}$ and $T_{H2}$). The chamber pressure has been varied between 50 and 80~bar. For that reason, all tested chamber pressures are close to or above the critical pressure of oxygen, at which LOX injection and combustion is transcritical. ROF between 2 and 7 has been achieved, while most load points are between ROF 4 and 6. During the operation of BKD at the test bench P8, the LOX injection temperature $T_{O2}$ is typically around 110~K. The hydrogen injection temperature $T_{H2}$ depends on the used H$_2$ storage system of the test bench \cite{Haberzettl} and is usually around 100~K. If a cryogenic storage tank is used, a lower injection temperature of 45 to 50~K can be achieved. For LOX/H$_2$ it has been reported that the hydrogen injection temperature has an impact on combustion stability \cite{Wanhainen1966,Nunome11,Yang}. For that reason, hydrogen temperature ramping tests have been performed in the past \cite{SGCEAS}. In this case $T_{H2}$ is varied between 45 and 135~K in several ramps. 

For the load point of 80~bar ROF 6 the thrust chamber achieves a theoretical thermal power of almost 100~MW and a thrust of about 24~kN. These numbers place BKD at the lower end of small upper stage engines.

The representative conditions in the thrust chamber with temperatures of up to 3600~K and high pressures limit the diagnostics for the instability investigations. A special measurement ring is placed between the injector head and the cylindrical combustion chamber segment, as indicated in Fig. \ref{fig_BKD}. At this location, the temperatures are still moderate due to the injection of cryogenic propellants, which allows the mounted sensors to survive the harsh conditions for several test runs. This measurement ring is extensively equipped with sensors, such as thermocouples or pressure sensors to measure the mean chamber pressure. All sensors are mounted in a common measurement plane, which is positioned 5.5~mm downstream of the injection plane. The main diagnostics for the stability analysis are water-cooled high-frequency piezoelectric pressure sensors. Eight Kistler type 6043A are flush-mounted in the ring with an even circumferential distribution, in order to measure the chamber pressure oscillations $p'$. The high-frequency pressure sensors have a measurement range set to $\pm30$~bar and a sampling rate of 100~kHz. An anti-aliasing filter was set at 30~kHz.

\begin{figure}
\includegraphics[width=8.5cm]{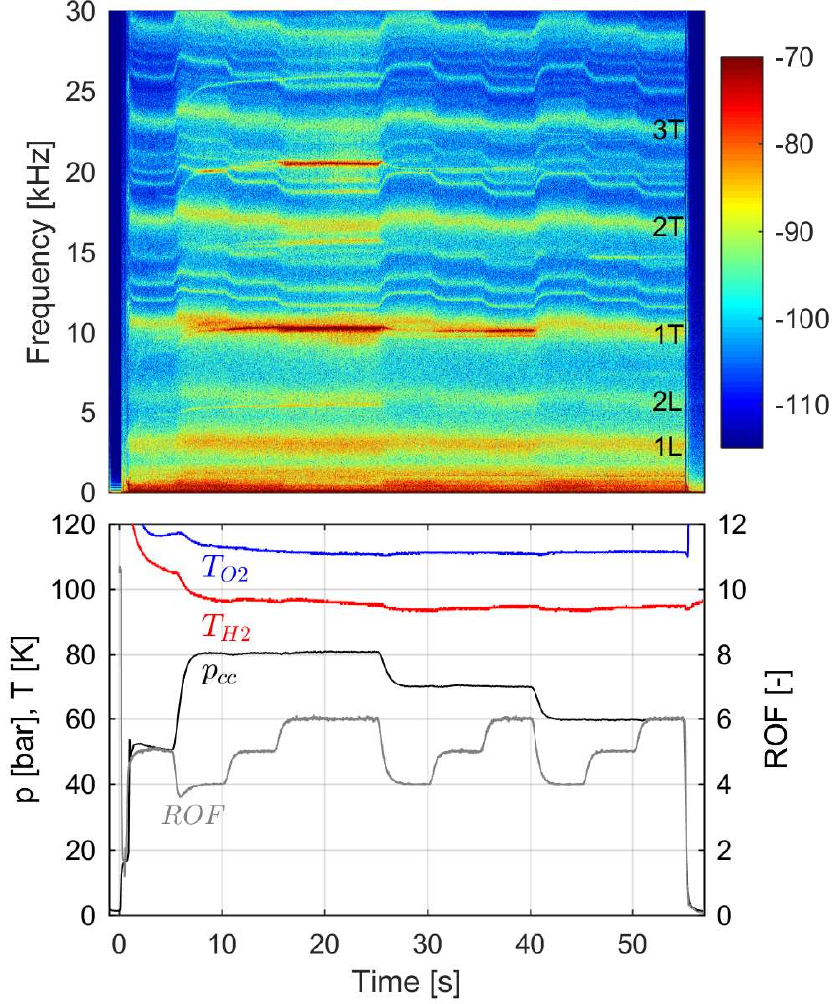}
\caption{\label{fig_Spectrogram_Type1} BKD test sequence and $p'$ spectrogram showing the type 1 instability \cite{SGJPP}.}
\end{figure}

Two different types of self-excited high-frequency combustion
instabilities have been detected, both of which are driven by injection coupling \cite{SGDiss,SGJPP,Hardi18_AA,Armbruster18_AA}.
A typical BKD test sequence, is shown in Fig. \ref{fig_Spectrogram_Type1}. A pressure oscillation spectrogram from inside the combustion chamber is also presented.

Stable and unstable operating conditions can be identified in the spectrogram. The strongest self-excited high-frequency combustion instabilities of the first tangential (1T) resonance mode at about 10~kHz were found for the 80 bar, ROF 6 load point between 16 and 25~s with a hydrogen injection temperature of $T_{H2} \approx 100$~K. The maximum peak-to-peak amplitudes during unstable combustion reached up to 16-35~bar but showed a large temporal variation.
In addition to the pressure oscillation data, Gröning et al. \cite{SGJPP,SGPropPhys} used fibre-optical probes to record fluctuating OH* radiation intensity of individual flames. These optical measurements are not included in the current study but helped to identify the instability driving mechanism in BKD.
The OH* fluctuations showed dominant frequencies corresponding to LOX post acoustic resonance frequencies, independent of chamber acoustics \cite{SGJPP,SGDiss}. For that reason, Gröning et al. described the coupling mechanism as injection-driven \cite{SGJPP}. The flame dynamics are modulated by the LOX post acoustics, and combustion instabilities emerge when the frequency of a chamber mode matches one of the longitudinal modes of the injectors. For the load points of $p_{cc}$ 80~bar ROF 6 this frequency interaction appears for the chamber 1T mode with the second longitudinal eigenmodes of the LOX posts \cite{SGJPP}. This mechanism was later confirmed with high-speed imaging of the flame dynamics in the thrust chamber using a 2D optical access window \cite{Armbruster19JPP}. Recent investigations indicated that an additional hydrodynamic effect in the LOX injectors may also play a role in the excitation of the LOX post eigenmodes and the amplification of the combustion instability \cite{Armbruster19JPP}. As was shown, the coupling mechanism of this type of instability seems well understood. Furthermore, the stability behavior shows reproducible characteristics. If the load point of 80~bar ROF 6 was approached in the test sequence, the chamber consistently showed an excitation of the 1T mode for the identical thrust chamber configuration \cite{SGJPP,SGDiss}.

\begin{figure}
\includegraphics[width=8.5cm]{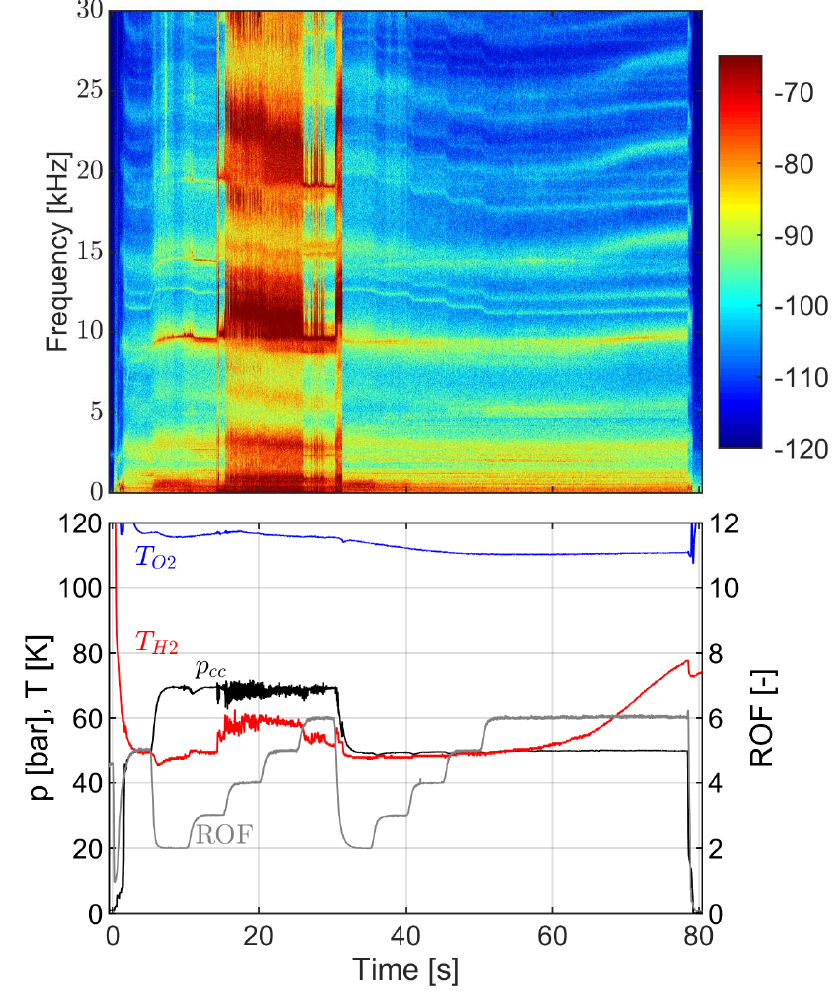}
\caption{\label{fig_Spectrogram_Type2} BKD test sequence and $p'$-spectrogram showing the type 2 instability \cite{Armbruster18_AA}}
\end{figure}

Beside the aforementioned LOX-coupled type instability, a second type with very high amplitudes (up to > 75~\% of $p_{cc}$) appears under rare circumstances \cite{Hardi18_AA,Armbruster18_AA}.
This type of instability appeared for different operating conditions and injector lengths. Most of the load points, which showed this type of instability, have a cold hydrogen injection temperature of about 45~K. Fig. \ref{fig_Spectrogram_Type2} shows a test sequence and $p’$-spectrogram of a run with cold hydrogen injection, which showed this type of instability.

As can be observed, the instability shows higher amplitudes compared to the type 1 instability, sometimes even exceeding the measurement range of the acoustic pressure sensors, and appears for different operating conditions. Furthermore, even for stationary operating conditions the high-amplitude pressure oscillations can appear and disappear spontaneously several times, as can, for example, be observed for the load point of 70~bar ROF 6 during the time window from 26-30~s. Other effects, which can be observed in Fig. \ref{fig_Spectrogram_Type2} are an increase of the hydrogen injection temperature during the instability and an apparent broadening of the acoustic spectrum due to switching of instability type and therefore frequency. Recent investigations of the acoustic field dynamics and the coupling mechanism indicated a three-way coupling with the chamber and both injection systems \cite{Armbruster18_AA}. At first, the instability seems to be initialized similarly to the type 1 instability, by a coupling of the LOX-posts with the chamber acoustics. However, for unknown reasons, the amplitudes grow significantly higher than that of the type 1 instability. The resulting strong transverse oscillations enhance the mixing of the injection propellants, which leads to short flames \cite{Armbruster18_AA}. Similar observations have been reported in other experiments \cite{Hardi14,Hakim} and simulations \cite{Hardi18_AA,BeinkeDiss,Hakim,Schmitt,Urbano16}.
Due to the shortened flames, the 1T mode frequency increases and allows coupling with the H$_2$ injectors \cite{Armbruster18_AA}.

The fact that this type of oscillation appears spontaneously during stationary operating conditions indicates that the system is being triggered by a nonlinear mechanism. Previous studies were not able to identify a triggering process although in gas turbines, triggering has been seen to be caused by background noise \cite{Zinn2005}, due to the non-normal behaviour of thermoacoustic systems \cite{Juniper2011}. For that reason, while being able to identify operating conditions with an increased risk for this type of instability, it is currently impossible to predict the onset of this instability. These characteristics make the prediction and analysis of the type 2 instability more complicated than the type 1 instability.

\section{Recurrence Quantification Analysis}
\label{sec_rqa}

In this section, we review the basics of recurrence quantification analysis (RQA) \cite{Marwan,Webber}. RQA was developed to study and compare recurrence plots which are used to visualize the recurrence behavior of the phase space trajectory of dynamical systems. By Takens' delay embedding theorem \cite{Takens}, we can reconstruct the dynamics of the rocket thrust chamber in an appropriate phase space from a single state variable such as the acoustic pressure $p'$ \cite{Juniper2018}. The univariate pressure time series data are converted into a set of delayed vectors:
\begin{equation}
    \bm{x}(t)=[p'(t),p'(t+\tau),p'(t+2\tau),\dots,p'(t+(d-1)\tau)],
\end{equation}
where $\tau$ and $d$ denote an appropriate time delay and embedding dimension respectively. The time delay $\tau$ can be estimated using the autocorrelation function or average mutual information \cite{Kasthuri}. The embedding dimension $d$ can be obtained using the false nearest neighbor method or Cao's method \cite{Cao}. After the reconstruction of a suitable phase space, the recurrence matrix is computed by the following formula: 
\begin{equation}
    R_{ij}=\Theta(\epsilon-\|\bm{x}(i)-\bm{x}(j)\|),
\end{equation}
where $\Theta$ is the Heaviside step function, $\epsilon$ is a predefined threshold, and $\|\bm{x}(i)-\bm{x}(j)\|$ is the Euclidean distance between the state points $\bm{x}(i)$ and $\bm{x}(j)$. Whenever a state point recurs in the predefined threshold, $R_{ij}$ is set to 1, otherwise $R_{ij}$ is set to 0. A recurrence plot is the two-dimensional arrangement of recurrence points, i.e. a visualisation of the recurrence matrix. Different patterns characterize different dynamics of the signal. Several statistical measures are used to quantify the structures present in a recurrence plot \cite{Marwan,Webber}. The recurrence rate (RR) measures the density of recurrence points:
\begin{equation}
    \text{RR}=\frac{1}{N^{2}}\sum_{i,j=1}^{N}R_{ij},
\end{equation}
where $N=n-(d-1)\tau$ is the number of state vectors in the reconstructed phase space. The determinism (DET) measures the percentage of recurrence points which form diagonal lines of minimal length $l_{min}$:
\begin{equation}
    \text{DET}=\frac{\sum_{l=l_{min}}^{N}l\,P(l)}{\sum_{l=1}^{N}l\,P(l)},
\end{equation}
where $P(l)$ is the probability distribution of diagonal lines having length $l$. The measure is called determinism because it is related to the predictability of the studied dynamical system. The laminarity (LAM) similarly quantifies the amount of recurrence points which form vertical lines:
\begin{equation}
    \text{LAM}=\frac{\sum_{v=v_{min}}^{N}v\,P(v)}{\sum_{v=1}^{N}v\,P(v)},
\end{equation}
where $P(v)$ is the probability distribution of vertical lines having length $v$, which have at least a length of $v_{min}$. The probability $p(l)$ that a diagonal line has exactly length $l$ can be estimated from the probability distribution $P(l)$:
\begin{equation}
    p(l)=\frac{P(l)}{\sum_{l=l_{min}}^{N}P(l)}.
\end{equation}
Using $p(l)$, one can further calculate the associated Shannon entropy (ENTR):
\begin{equation}
    \text{ENTR}=-\sum_{l=l_{min}}^{N}p(l)\ln{p(l)},
\end{equation}
which reflects the complexity of the deterministic structure in the dynamical system. An additional measure is the ratio of the determinism to the recurrence rate, i.e. RATIO=DET/RR.

\section{Support Vector Machines}
\label{sec_svm}

In this section, we briefly recall the basics of support vector machines (SVMs) \cite{Vapnik,Bishop}. SVMs are machine learning models with associated (supervised) learning algorithms, which are used for classification and regression tasks. In binary classification, one is given a training dataset of $n$ points of the form
\begin{equation}
    (\bm{x}_{1},y_{1}),\dots,(\bm{x}_{n},y_{n}),
\end{equation}
with $d$-dimensional patterns $\bm{x}_{i}$ and associated class labels $y_{i}\in\{\pm 1\}$, and tries to estimate a function $f$ such that $f$ will correctly classify new examples $(\bm{x},y)$, i.e. $f(\bm{x}) = y$ for points which were generated from the same underlying probability distribution as the training data. SVMs assume that the decision function $f$ is based on the class of hyperplanes
\begin{equation}
    (\bm{w}\cdot\bm{x})+b=0,
\end{equation}
where $\bm{w}\in\mathbb{R}^{d}$ and $b\in\mathbb{R}$. The associated function $f$ is of the form
\begin{equation}
    f(\bm{x})=\text{sign}((\bm{w}\cdot\bm{x})+b).
\end{equation}

The goal is to find the hyperplane with the maximal margin of separation between the two classes. In general, the larger the margin, the lower the generalization error of the classifier. If the training data is linearly separable, this leads to the following optimization problem:
\begin{equation}
    \min\|\bm{w}\|^{2}\quad\text{subject to}\quad y_{i}((\bm{w}\cdot\bm{x}_{i})+b)\leq 1,
\end{equation}
for $i=1,\dots,n$. The above is an optimization problem with a convex quadratic objective and only linear constraints. It can be solved using quadratic programming techniques. Recent methods for finding the SVM classifier also include sub-gradient descent and coordinate descent algorithms. The optimal hyperplane is completely determined by those $\bm{x}_{i}$ that lie nearest to it. These data points are called support vectors. It follows that $\bm{w}$ is given by a linear combination of the support vectors,
\begin{equation}
    \bm{w}=\sum_{i}\nu_{i}\bm{x}_{i},
\end{equation}
and the decision function can be written as
\begin{equation}
    f(\bm{x})=\text{sign}(\sum_{i}\nu_{i}(\bm{x}\cdot\bm{x}_{i})+b).
\end{equation}

To extend SVMs to cases in which the data are not linearly separable, one introduces slack variables $\xi_{i}$ to allow certain constraints to be violated. The new optimization problem can be expressed as
\begin{equation}
    \min\|\bm{w}\|^{2}+C\sum_{i}\xi_{i}\quad\text{subject to}\quad y_{i}((\bm{w}\cdot\bm{x}_{i})+b)\leq 1-\xi_{i}
\end{equation}
and $\xi_{i}\leq 0$, for all $i$. Nonlinear classifiers can be constructed by applying the kernel trick below. The resulting algorithm is formally similar, except that every dot product is replaced by a nonlinear kernel function
\begin{equation}
    k(\bm{x},\bm{y}):=(\bm{\Phi}(\bm{x})\cdot\bm{\Phi}(\bm{y})),
\end{equation}
where $\bm{\Phi}$ is a nonlinear map from $\mathbb{R}^{d}$ into a feature space $\mathbb{F}$,  $\bm{\Phi}:\mathbb{R}^{d}\to\mathbb{F}$. Although the algorithm fits a hyperplane in the transformed feature space, the decision function may be nonlinear in the original input space,
\begin{equation}
    f(\bm{x})=\text{sign}(\sum_{i}\nu_{i}k(\bm{x},\bm{x}_{i})+b).
\end{equation}

The hyperparameters consist of the soft margin constant, $C$, and parameters on which the kernel function may depend, e.g. the width of a Gaussian kernel.

\section{Early Warning Signal Construction}
\label{sec_method}

In this section, we present the proposed construction of a suitable early warning signal. Clearly, a suitable warning signal could be constructed using a combustion noise feature that increases or decreases monotonically before an instability and a universal threshold, which separates stable data points depending on whether they belong to the onset of instability or not. The comparison of the current value with the threshold could be used as an early warning signal. In general, this will not be possible and the classification using a simple threshold value for a certain feature will not correspond to a division into data points that belong to the onset of instability or not. Naively, one could think that the magnitude of the amplitude spectrum of the pressure signal at the dominant frequency constitutes a suitable measure by setting an appropriate threshold. A closer look shows that this measure is not well suited, because of the dependency on operating conditions, propellant combinations, and combustion chamber geometries \cite{Kasthuri}. Furthermore, the root mean square (RMS) of the pressure signal does not increase in a monotonic way as the system approaches a combustion instability \cite{Kasthuri}. Thus, for such features, it is not possible to define universal thresholds to detect the onset of thermoacoustic combustion instability.

To overcome the shortcomings of conventional measures, different measures from nonlinear time series analysis have been introduced. Certain nonlinear combustion noise features are more independent of operating conditions and the exact combustion chamber geometries. RQA measures quantify recurrence behaviors, which are different for stable combustion and combustion instability \cite{Juniper2018}. The pressure signals in the unstable (limit-cycle) regime possess a deterministic periodic nature, while the stable regime is distinguished by a noisy or chaotic nature. Furthermore, the transient phase is characterized by a strong change in these measures. Because of this, it makes sense not only to look at the current values, which describe the present recurrence behavior, but also to investigate their variation. Trend estimation techniques can be used to determine whether time series data exhibit an increasing or decreasing trend. The linear fit for different sample windows is one of the simplest methods. It is expected that the consideration of trend behavior should increase predictability. A combination of several combustion noise features reduces the influence of outliers and usually makes forecasts more stable. To find the optimal combination and decision criterion respectively, one can use data-driven machine learning algorithms. 

Thus, we apply the following procedure:
\begin{itemize}
    \item calculate RQA measures and estimate their trends using suitable sample windows
    \item train an SVM to learn the associated magnitudes and trends in the data that belong to different regimes
    \item use the class prediction of the trained SVM as an online early warning signal
\end{itemize}

Compared to the use of only one combustion noise feature, the proposed method is expected to generate fewer false alarms and has the potential to work with other combustion chamber geometries due to the use of more universal combustion noise features. The central challenge is that the machine learning model may overfit during training by memorizing properties of the training data that do not work well on unseen data \cite{Bishop}. The capability to perform well on unseen input data is called generalization and can be estimated by using only a fraction of the available data for training. The remaining portion is used to select the optimal hyperparameter combination.

\section{Test Case}
\label{sec_test_case}

For the evaluation, we use a data set of 10 typical BKD tests. Table \ref{tab:table4} summarizes the main characteristics of each run.

\begin{table}
\caption{\label{tab:table4}Overview of used BKD test runs. Fig. \ref{fig_runs_1} and Fig. \ref{fig_runs_2} in the appendix show the growth of the normalized peak-to-peak amplitudes of the pressure oscillations and the associated test sequences.}
\begin{ruledtabular}
\begin{tabular}{lccc}
Run&Propellant state\footnote{GH (gaseous H2) and LH (liquid H2) refer to different fuel interfaces of the test bench. At the LH interface the hydrogen is stored as a cryogenic fluid, while at the GH interface it is stored in a high-pressure tank under ambient conditions and a heat exchanger is to used to further cool it down. For both cases, hydrogen is injected as a supercritical fluid, but the injection temperatures vary. The LH interface leads to an injection temperature around 45~K. The GH interface results in an injection temperature around 100~K.}&Post length\footnote{Most experiments used a LOX-post with a length of 68~mm. Additional tests with modified lengths, e.g. 64~mm, were also carried out.} $[mm]$&Instability type\\
\hline
1 & GH & 68 & 1\\
2 & GH & 68 & 1\\
3 & GH & 68 & 1\\
4 & GH & 68 & 1\\
5 & GH & 68 & 1\\
6 & LH & 68 & 1,2\\
7 & GH & 64 & 2,1\\
8 & GH & 68 & 2,1,1\\
9 & LH & 68 & 1,2\\
10 & LH & 68 & 1,2\\
\end{tabular}
\end{ruledtabular}
\end{table}

In total there are 16 instabilities present in the set of time series. The threshold of a data point to be classified as type 1 instability is set to a peak-to-peak amplitude of 6.25~\% with respect to the mean chamber pressure. This threshold marks the transition to limit cycle oscillations in BKD. A threshold of 5~\% would be violated by single spikes, which are not part of a lasting instability. In terms of type 2 instabilities, a threshold of 20.0~\% is chosen due to their large amplitudes. Fig. \ref{fig_runs_1} and Fig. \ref{fig_runs_2} in the appendix show the growth of the normalized peak-to-peak amplitudes and the associated test sequences. The training and validation data set is given by the BKD runs $[1,2,4,5,6,9]$. The test data set is given by the remaining BKD runs $[3,7,8,10]$.

First, for each time series, i.e. pressure signal of a BKD run, we calculate the points in time when an instability begins and when an instability ends. We define the start of an unstable combustion process by the fact that the peak-to-peak amplitude increases above 6.25~\% of the mean chamber pressure. An instability ends when the value drops below 6.25~\% of the mean chamber pressure again and stays there for at least 500~ms. The second condition causes a fast sequence of stable and unstable phases to be classified as unstable. This is desired because we intend to predict the first occurrence of strong pressure fluctuations.

Second, the following RQA measures are calculated using a sliding window approach: RR, DET, LAM, ENTR, and RATIO. The calculation of the quantities for time $t$ uses the values of the dynamic pressure signal from the interval $[t-200~ms,\,t]$. For each measure, we also estimate the time-dependent slope of the trend using a linear fit and windows of the form  $[t-100~ms,\,t]$. Further details of the calculation are described in the appendix. In total, we obtain ten derived combustion noise features.

To characterize the transient regime, the data points which belong to the stable phase are divided into two sets. The first set contains all stable data points whose points in time are at least 200~ms away from the occurrence of instability, while the second set includes all stable data points which are not in the first set. An SVM is used to solve this binary classification problem. We use a random search to find the best hyperparameters. To compare different hyperparameter combinations, we use cross-validation to assess the performance of the SVM measured by the F-score (harmonic mean of precision and recall). Data from one run are used to evaluate the prediction accuracy, while data from the other runs are used to train the SVM for a given hyperparameter combination. This procedure is repeated, such that data from every run are used for evaluation once. Finally, the performance is averaged. The hyperparameters which belong to the best average performance are used to train the final SVM using the complete training and validation data set. Finally, the prediction of this SVM is investigated using the test data set.

\section{Results}
\label{sec_results}

This section presents the results of the presented data-driven early warning signal method for the BKD test case. The performance of the classifier reflects the capability to successfully predict if the system will develop a thermoacoustic instability within the next 200\,ms using dynamic pressure sensor data. In other words, we quantify if the model can detect a divergence from the stable regime. Online prediction with such a time window provides a realistic lead-time for taking appropriate control actions.

\begin{figure}
\includegraphics{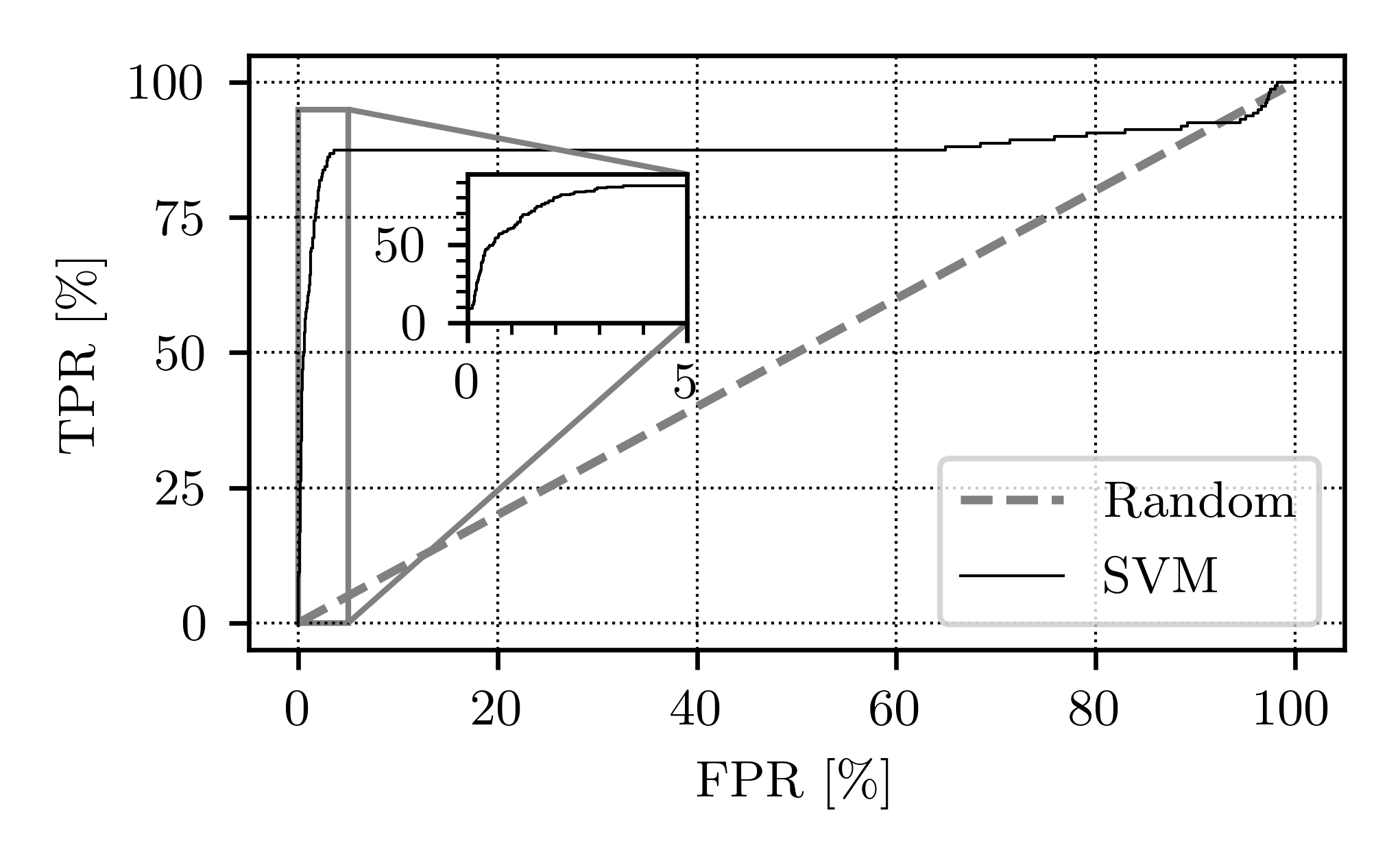}
\caption{\label{fig_ROC_curve}Receiver operating characteristic (ROC) curve of the test data set for the trained SVM classifier. Points above the diagonal represent good classification results (better than random).}
\end{figure}

For a classification problem with imbalanced classes, i.e. where the number of samples in the classes differs greatly as in our case, it is crucial to use suitable performance measures. The true-positive rate (TPR) is given by the number of true positives, i.e. the number of data points correctly labeled as belonging to the positive class (transient regime in our case), divided by the total number of data points that belong to the positive class. The false-positive rate (FPR) is given by the number of false positives, i.e. the number of data points incorrectly labeled as belonging to the positive class, divided by the total number of data points that belong to the negative class (far from instability in our case). A high FPR is equivalent to a large number of false alarms, i.e. a warning signal wrongly predicting a thermoacoustic instability. Thus, for suitable early warning signals, it is extremely important to exhibit a small FPR. Table \ref{tab_svm} shows the TPRs associated with the trained SVM for different FPRs. For an FPR of 0.5~\% the TPR is 49~\%. The TPR increases to 61~\% and 80~\% for an FPR of 1~\% and 2~\% respectively. A receiver operating characteristic (ROC) curve, can be used to illustrate the diagnostic ability of a binary classifier as its discrimination threshold is varied. It is created by plotting the TPR against the FPR at various threshold settings. Fig. \ref{fig_ROC_curve} displays the ROC curve for the SVM. One can see that the TPR remains limited to about 80~\% even for large FPRs.

\begin{figure*}
\includegraphics{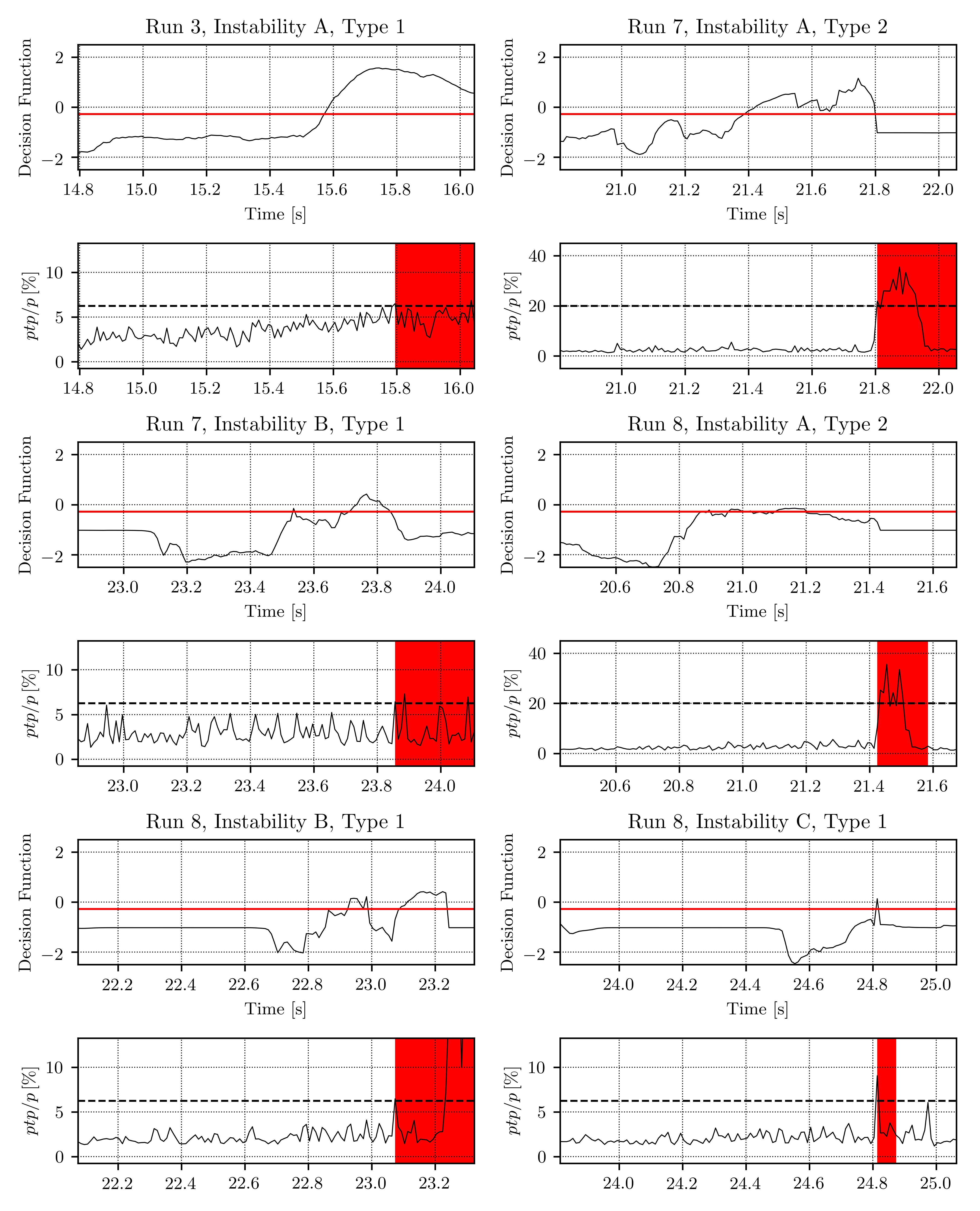}
\caption{\label{fig_decision_function_1}Predicted decision function values and the threshold corresponding to an FPR of 1~\% near the onset of combustion instability for the first part of the test data set. For evaluation of the early warning signal, the normalized peak-to-peak amplitude of the pressure oscillations is also shown. The start of combustion instability is defined by the condition that the normalized peak-to-peak amplitude increases above 6.25~\% (type 1) and 20.0~\% (type 2) respectively. An instability ends when the value drops below 6.25~\% again and stays there for at least 500~ms. The area of instability is marked red.}
\end{figure*}

\begin{figure*}
\includegraphics{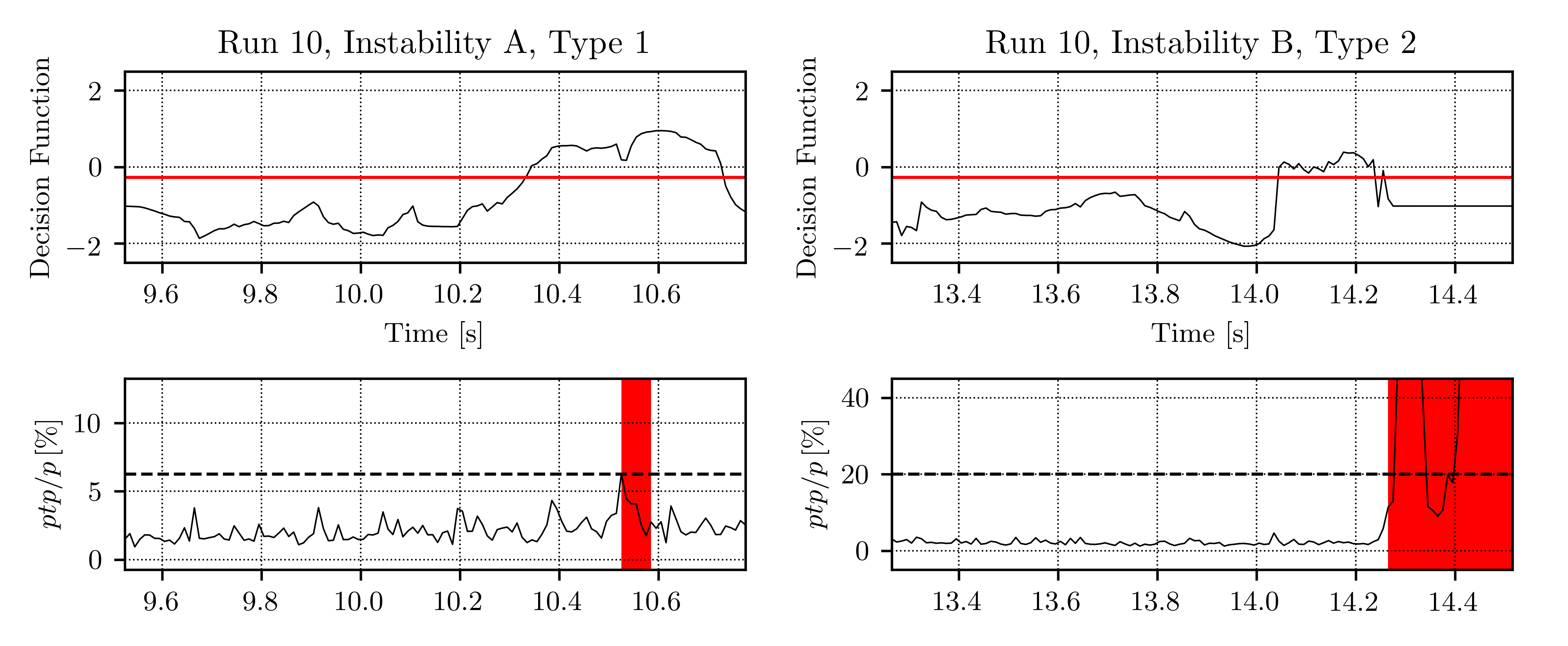}
\caption{\label{fig_decision_function_2}Predicted decision function values and the threshold corresponding to an FPR of 1~\% near the onset of combustion instability for the second part of the test data set. For evaluation of the early warning signal, the normalized peak-to-peak amplitude of the pressure oscillations is also shown. The start of combustion instability is defined by the condition that the normalized peak-to-peak amplitude increases above 6.25~\% (type 1) and 20.0~\% (type 2) respectively. An instability ends when the value drops below 6.25~\% again and stays there for at least 500~ms. The area of instability is marked red.}
\end{figure*}

By looking at the decision function of the SVM for all instabilities in the test set, the responsible reasons become apparent. The decision function and the threshold associated with an FPR of 1~\% is shown in Fig. \ref{fig_decision_function_1} and Fig. \ref{fig_decision_function_2}.

\begin{table}
\caption{\label{tab_svm}True-positive rate (TPR) of the SVM classifier for different false-positive rates (FPRs). The FPRs are realized by varying the decision threshold.}
\begin{ruledtabular}
\begin{tabular}{lccc}
 &TPR(FPR=0.5\,\%)&TPR(FPR=1\,\%)&TPR(FPR=2\,\%)\\
\hline
SVM & 49\,\% & 61\,\% & 80\,\% \\
\end{tabular}
\end{ruledtabular}
\end{table}

The instability in run 3 is predicted around 200\,ms before occurrence. The type 2 instability in run 7 is predicted within 200\,ms, but the value of the decision function exceeds the threshold 400\,ms before the instability begins. Thus, there are false positives. The transition to the type 1 instability is detected as intended. Nevertheless, the performance for this run is very convincing, because run 7 was carried out with a modified injector geometry and there are no time series for this setup in the training data. The fact that the prediction still performs so well indicates that the early warning signal could also work with modified injection systems. The performance in run 8 is not convincing. Only the second instability, which is of type 1, is satisfactorily predicted. At the first instability (type 2) the warning signal flickers but goes out before the instability. The third instability of type 1 is detected too late. The performance at the last run in the test set is again very convincing and the early warning signal works as intended. This is astonishing because the time series belongs to an experiment with a reduced injection temperature of the hydrogen. 

All in all, in 6 of 8 instabilities (type 1 and type 2) the constructed early warning signal works in a satisfying way. This leads to a TPR of about 60~\%. The TPR can be improved by reducing the threshold, but this also increases the number of false alarms. To predict the instabilities that are present at run 8 in time, the threshold must be significantly reduced. This leads to FPRs that are no longer acceptable.

\begin{figure}
\includegraphics{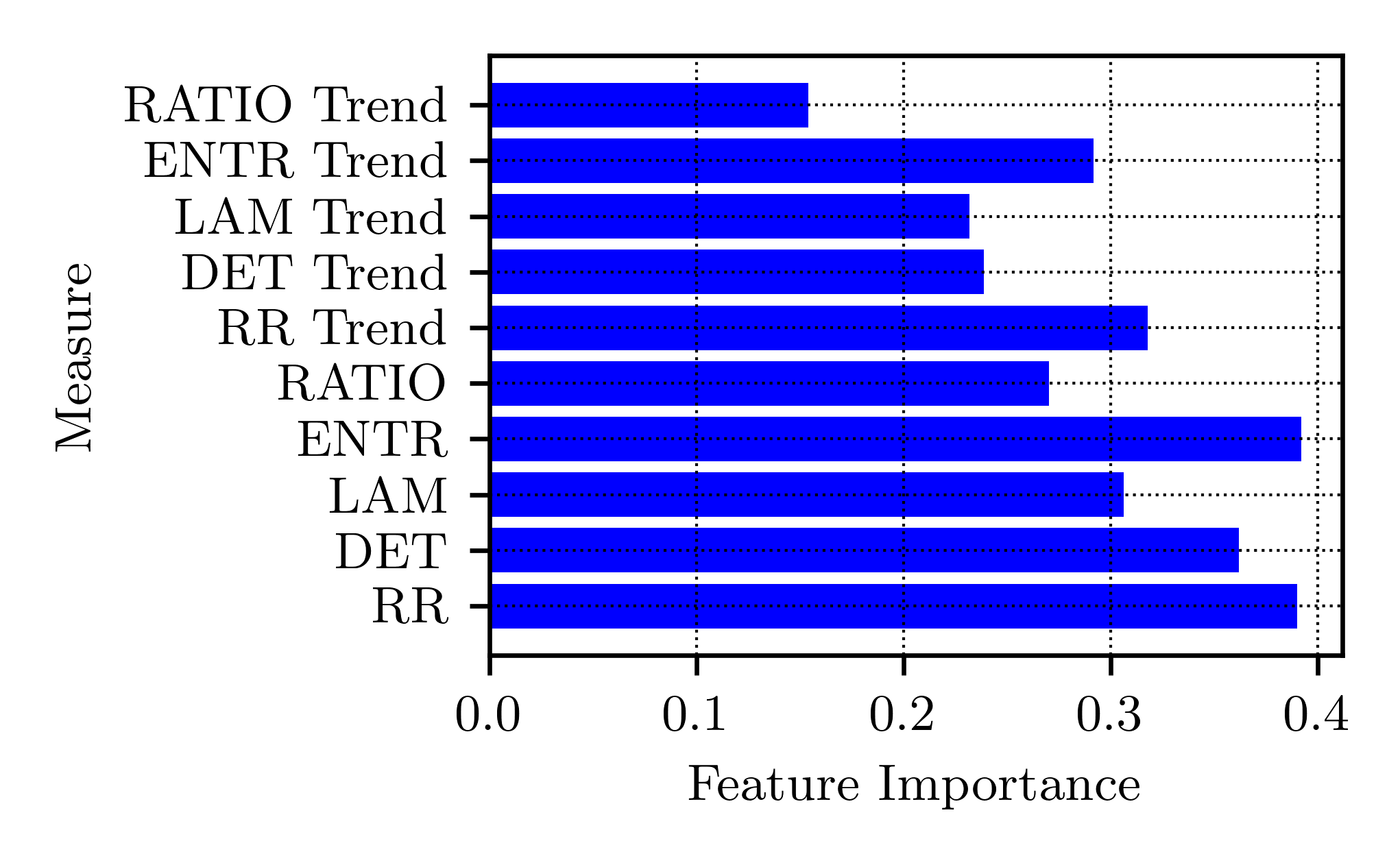}
\caption{\label{fig_feature_importance}Permutation feature importance plot. Feature importances are measured through the decrease of the F-score on the test data set.}
\end{figure}

To estimate the relative importance of input features, one can use the permutation feature importance method. Permutation feature importance measures the decrease in the prediction accuracy of the model after permuting the feature. A feature is more important if shuffling its values decreases the performance to a greater extent. Fig. \ref{fig_feature_importance} shows the feature importances measured through the features' influence on the F-score. The most important input features are given by the recurrence rate RR and the Shannon entropy ENTR.

For comparison, we also calculate the TPRs and FPRs, which result from the use of a single measure. By varying the threshold, the TPR can be determined for different FPRs. The results are shown in Table \ref{tab_comparison}. For an FPR of 1~\% the TPRs are less than or equal to 40~\%. Our proposed method leads to a TPR of 61~\%.

\begin{table}
\caption{\label{tab_comparison}True-positive rates (TPRs) associated with single RQA measures for different false-positive rates (FPRs). The FPRs are realized by varying the decision thresholds.}
\begin{ruledtabular}
\begin{tabular}{lccc}
Measure&TPR(FPR=0.5\,\%)&TPR(FPR=1\,\%)&TPR(FPR=2\,\%)\\
\hline
RR & 6\,\% & 32\,\% & 59\,\% \\
DET & 14\,\% & 35\,\% & 46\,\% \\
LAM & 2\,\% & 40\,\% & 71\,\% \\
ENTR & 32\,\% & 36\,\% & 38\,\% \\
RATIO & 2\,\% & 33\,\% & 62\,\% \\
\end{tabular}
\end{ruledtabular}
\end{table}

\begin{figure}
\includegraphics{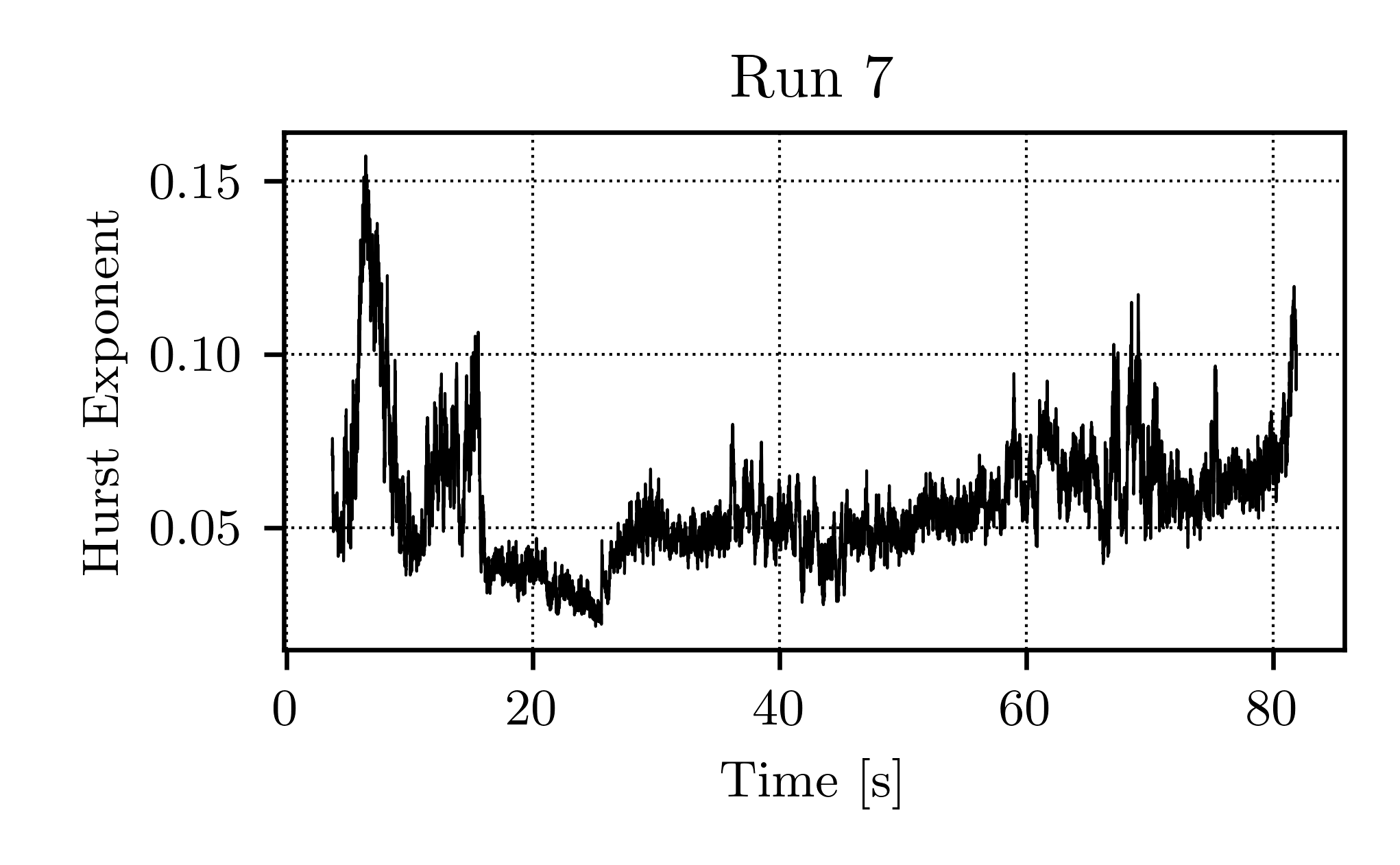}
\caption{\label{fig_hurst}Hurst exponent for run 7.}
\end{figure}

Another measure that can be calculated from the combustion noise is the Hurst exponent $H$. There is a loss of multifractality in combustion noise as combustors progress towards combustion instability, which is reflected in a decline of $H$ \cite{Nair2014,Juniper2018}. We compute $H$ by calculating the slope of the detrended fluctuations for logarithmically spaced window sizes ranging from 50\,000 to 70\,000 data points (corresponding to 5 to 7 cycles of the 10\,kHz instability). In the paper of Nair et al \cite{Nair2014}, it is suggested to use 2 to 4 cycles, but the slope is changing a lot in this range. Thus, we increase the number of cycles to get the asymptotic slope of the fluctuations. Although $H$ drops to a very low value when the instability is present, the performance as an early warning signal is not convincing. Table \ref{tab_hurst} shows the prediction accuracy using an optimal threshold for the test data set. When using 2 to 4 cycles for the computation of $H$, the prediction accuracy is even lower. The main reason for this is that $H$ fluctuates strongly and therefore generates many false alarms. As an example, Fig. \ref{fig_hurst} displays the time development of $H$ for run 7.

\begin{table}
\caption{\label{tab_hurst}True-positive rate (TPRs) associated with the Hurst exponent $H$ for different false-positive rates (FPRs). The FPRs are realized by varying the decision thresholds.}
\begin{ruledtabular}
\begin{tabular}{lccc}
&TPR(FPR=0.5\,\%)&TPR(FPR=1\,\%)&TPR(FPR=2\,\%)\\
\hline
$H$ & 11\,\% & 24\,\% & 27\,\% \\
\end{tabular}
\end{ruledtabular}
\end{table}

\section{Conclusion and Outlook}
\label{sec_conclusion}

In this paper, we presented the data-driven construction of a warning signal for the early detection of thermoacoustic instabilities in rocket thrust chambers. Using time-series data from experimental tests performed with the cryogenic rocket thrust chamber BKD, we calculated time-dependent RQA measures and their variation. Then we applied machine learning to obtain the characteristic properties of the transition phase. After the training and tuning phase, we evaluated the method in a quantitative way and compared it with other early warning indicators.

The constructed early warning signal achieved the best performance and was able to predict 6 of 8 thermoacoustic instabilities (type 1 and type 2) 200\,ms in advance with a low probability of false alarms. Furthermore, the results indicate that there is a good transferability to other injection systems. There might be a good transferability to other combustion chamber geometries. With this approach, high-frequency combustion instabilities can be successfully predicted under representative conditions. This is an important step towards the active control of combustion instabilities in the future.

There are essential questions that should be investigated in the future. Besides the systematic examination of the generalization ability, the influence of different noise levels should be examined. Since many research thrust chambers like the BKD are equipped with diverse sensors, methods should also be investigated, which fuse data from multiple sensors.

Information contained in the pressure signal is lost in the calculation of the RQA measures. Therefore machine learning methods should be investigated which derive suitable features directly from the data, e.g. deep neural networks.

Machine learning techniques can perform poorly when given inputs dissimilar to those seen during training, so-called out-of-distribution data. This can be problematic for safety-critical systems like rocket engines, particularly since our algorithms are trained on data sets whose size is necessarily limited by the resource-intensive nature of the experiments. We are therefore exploring Bayesian neural networks as a tool to quantify predictive uncertainties and make our predictions robust to overconfident extrapolations.

\begin{acknowledgments}

The authors would like to thank the crew of the P8 test bench and also Stefan Gröning for preparation and performing the test runs. Furthermore, it is a pleasure to thank Kai Dresia and Christoph Räth for many useful discussions.

\end{acknowledgments}

\appendix

\section{Calculation Details}

\begin{figure}
\includegraphics{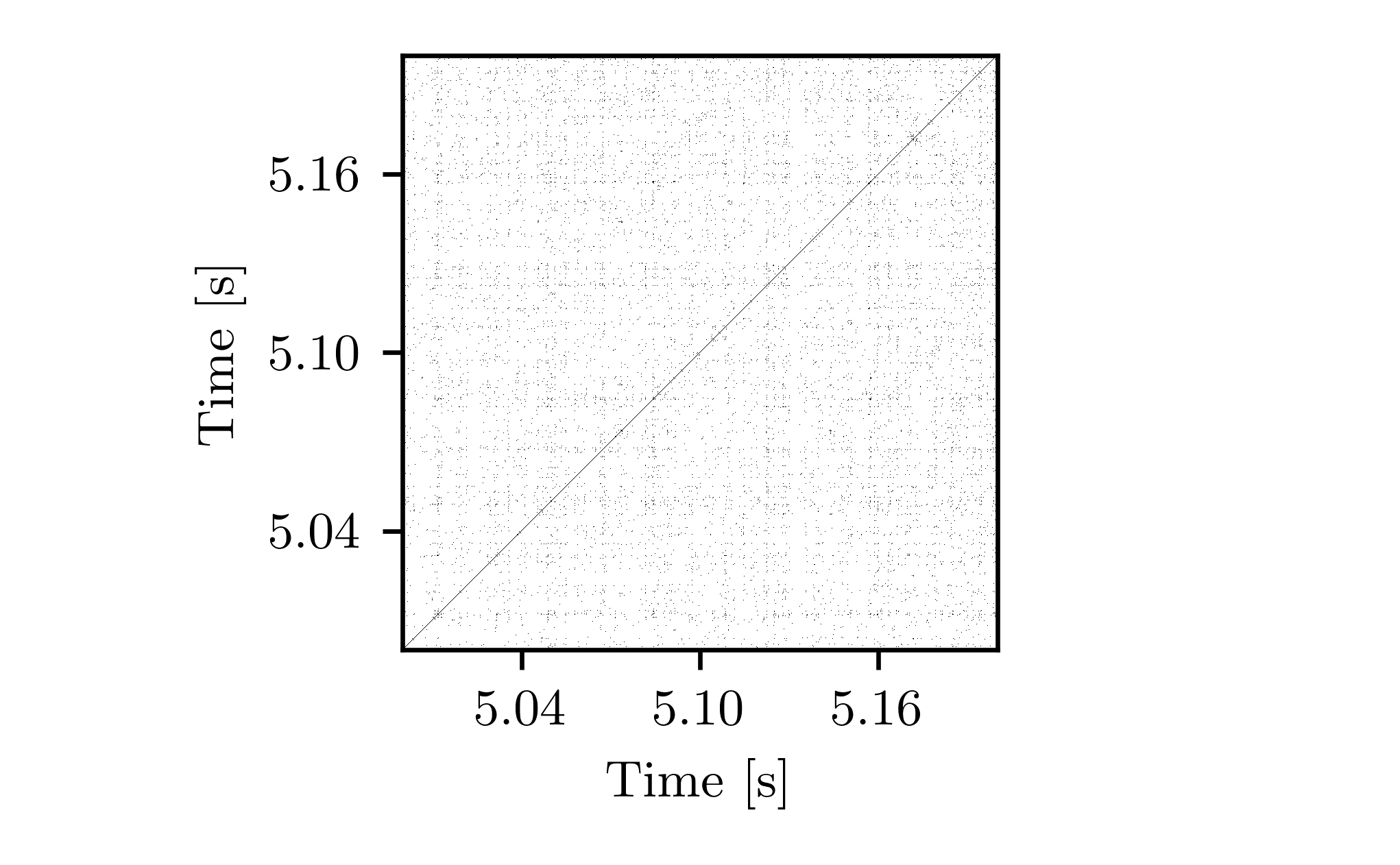}
\caption{\label{fig_rm_stable}Recurrence plot for the typical dynamics of stable operation.}
\end{figure}

\begin{figure}
\includegraphics{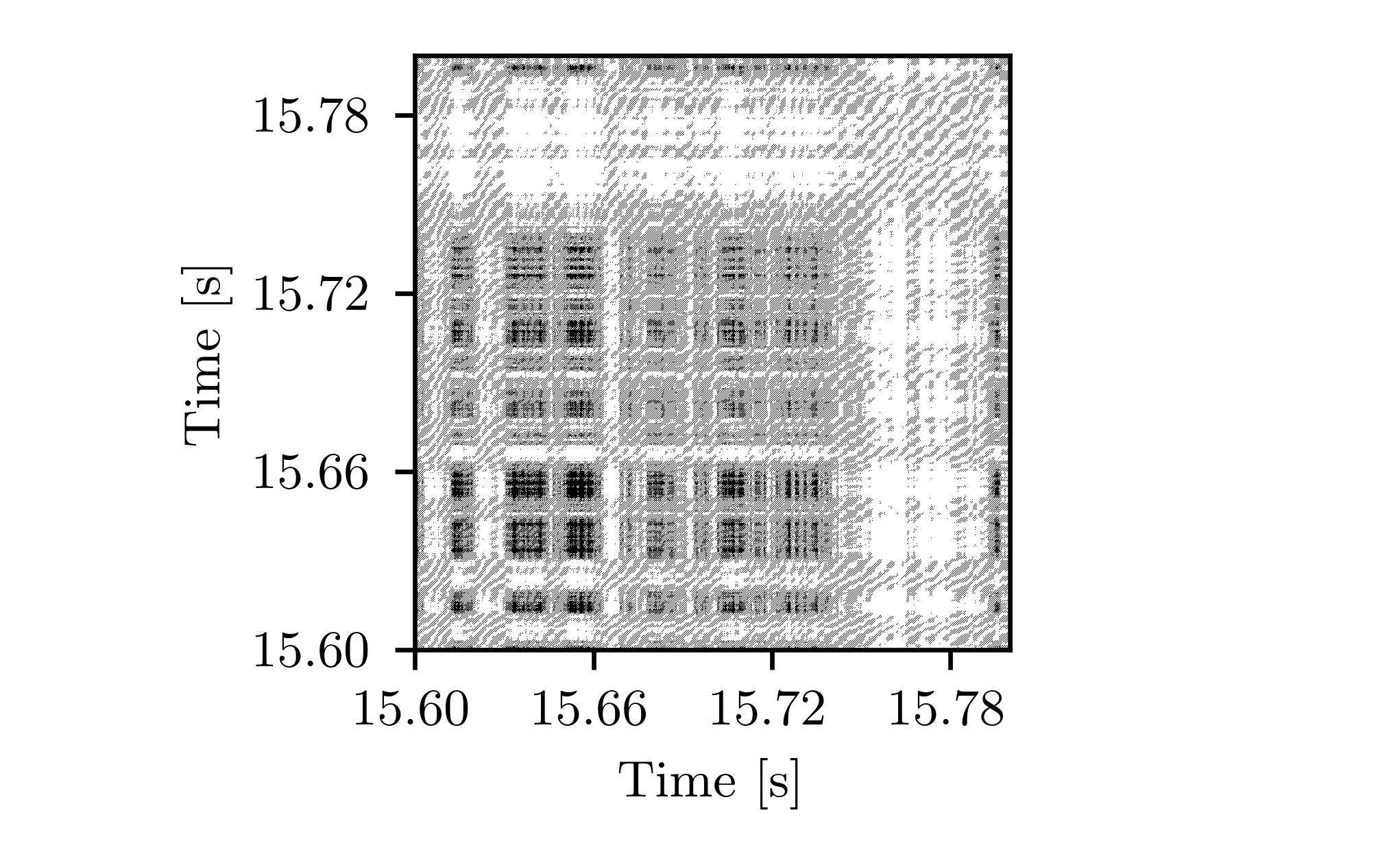}
\caption{\label{fig_rm_before}Recurrence plot for the typical dynamics of a transition to instability.}
\end{figure}

The basis of all calculations is the signal of a single dynamic pressure sensor. A high-pass filter is used to remove non-acoustic contributions. For reconstructing the phase space that is traced by the pressure oscillation, we estimate the optimum time delay and embedding dimension using the autocorrelation function \cite{Nayfeh} and Cao's method \cite{Cao} respectively. The optimum time delay corresponds to the first zero crossing of the autocorrelation function and is given by 0.02~ms. We chose a constant embedding dimension of 15 for all time series. For details related to the method of Cao to identify the minimum embedding dimension, the reader is referred to the paper of Cao \cite{Cao}. The time-dependent recurrence plots use the signal values from the last 200~ms in a sliding window manner. The threshold is set to 3 distance units after rescaling the input. Fig. \ref{fig_rm_stable} and Fig. \ref{fig_rm_before} show two exemplary recurrence plots. The pattern that is visible in Fig. \ref{fig_rm_stable} is typical for a stable combustion process. The pattern that is shown in Fig. \ref{fig_rm_before} characterizes a transition to instability.

By using recurrence plots analogous to the shown ones, RQA measures are calculated using the pyunicorn Python package \cite{Donges}. For trend estimation, we employ values of the last 100~ms in a sliding window manner and a simple linear fit. Then, all input features are standardized by removing the mean and scaling to unit variance. We use the standard scikit-learn \cite{scikit-learn} SVM implementation for the binary classification and adjust the class weights inversely proportional to class frequencies in the input data. The optimal hyperparameters turn out to be a soft margin constant $C=0.195$ and a radial basis function kernel coefficient $\gamma=0.918$.

\section{BKD Runs}

\begin{figure*}
\includegraphics{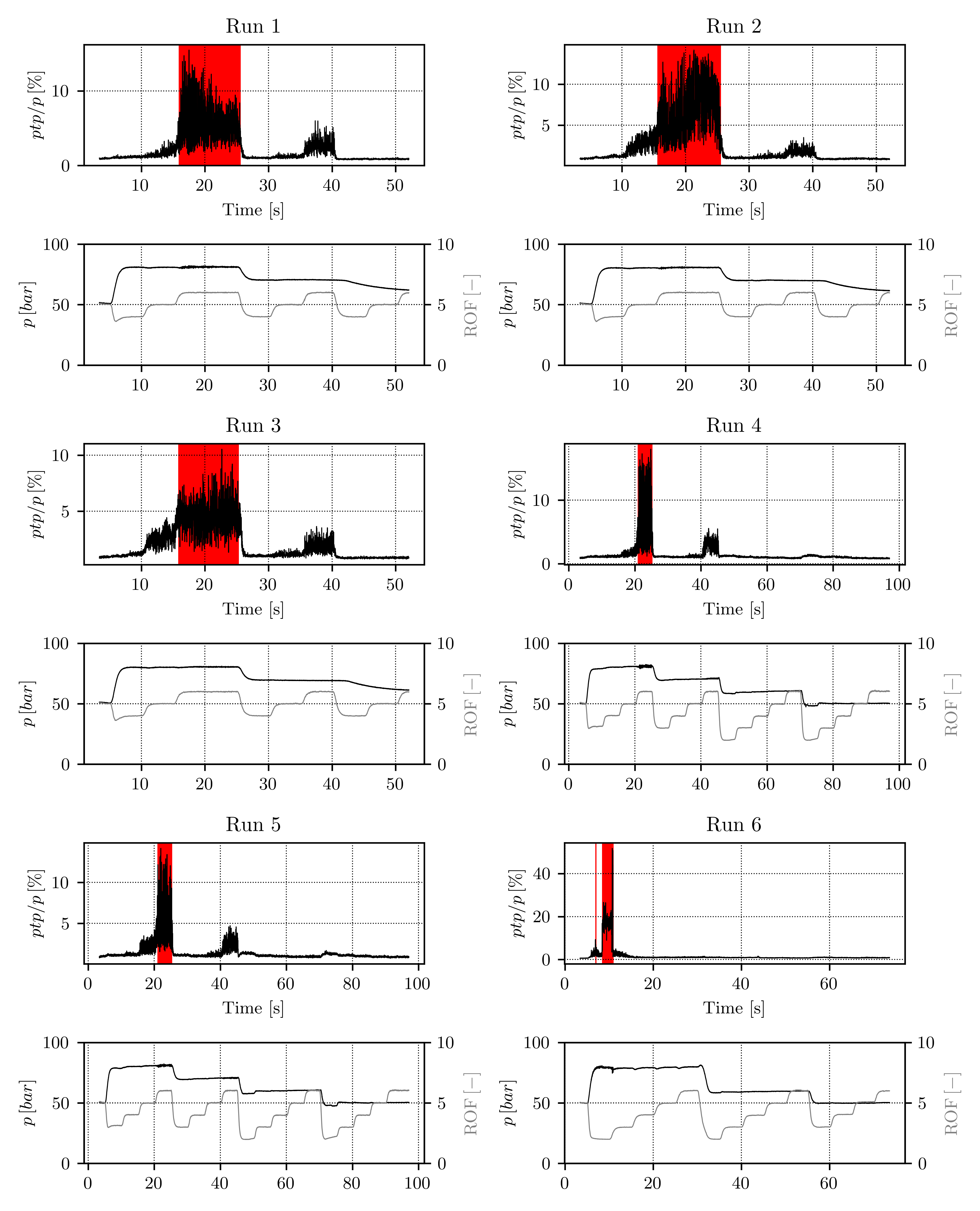}
\caption{\label{fig_runs_1}Normalized peak-to-peak amplitudes of the pressure oscillations and the associated test sequences for the BKD runs used in the current study. The area of instability is marked red.}
\end{figure*}

\begin{figure*}
\includegraphics{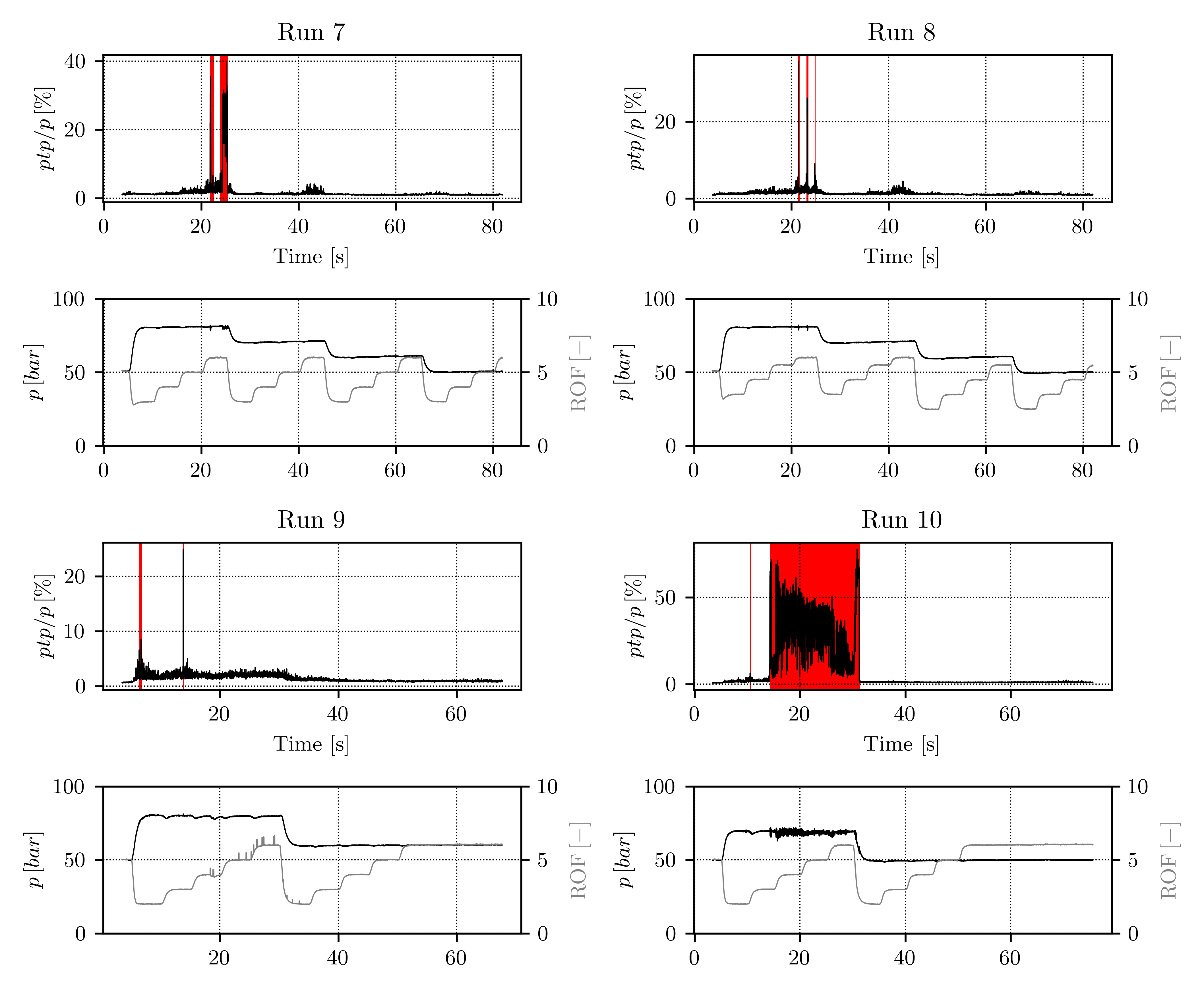}
\caption{\label{fig_runs_2}Normalized peak-to-peak amplitudes of the pressure oscillations and the associated test sequences for the BKD runs used in the current study. The area of instability is marked red.}
\end{figure*}

Fig. \ref{fig_runs_1} and Fig. \ref{fig_runs_2} show the normalized peak-to-peak amplitudes and the associated test sequences for all BKD runs used in the current study. 

\bibliography{DissBib}

\end{document}